\documentclass[%
 reprint,
 amsmath,amssymb,
 aps,
]{revtex4-1}

\usepackage{graphicx}
\usepackage{dcolumn}
\def\sp{\vspace{.1in}}
\def\no{{\noindent}}
\usepackage{bm}

\begin{document}

\preprint{APS/123-QED}
\title{Exact Geometries from Boundary Gravity}

\author{Rohit K. Gupta, Supriya Kar, R. Nitish and Monika Verma}

\affiliation{Department of Physics and Astrophysics\\ University of Delhi, New Delhi 110007, India}



\date{\today}

\begin{abstract}
 We show that the extremal Reissner-Nordstr\"{o}m type multi black holes in an emergent scenario are exact in General Relativity. It is shown that an axion in the bulk together with a geometric torsion ensure the required  energy-momentum to source the $(3$$+$$1)$ geometry in the Einstein tensor. Analysis reveals a significant role of dark energy to the curved space-time.            
\end{abstract}

\maketitle


\section{\label{sec:level1}Introduction}
Einstein gravity elegantly ensures the deformation geometries of space-time underlying the Riemann curvature tensor ${\cal R}_{\mu\nu\lambda\rho}$. It is a classical theory of metric $g_{\mu\nu}$ field and delves with the continuum space-time.  Furthermore the ${\cal R}_{\mu\nu\lambda\rho}$ is reducible and is re-expressed with three irreducible curvature tensors, $i.e.$ Conformal-Weyl $C_{\mu\nu\lambda\rho}$, trace-less Ricci $S_{\mu\nu}$ and Ricci scalar ${\cal R}$. Thus the curved  space-time in Einstein gravity is contained in one or more of the three irreducible curvatures.  However the Einstein gravity was elegantly 
described by the Einstein-Hilbert action which is expressed only in terms of the scalar ${\cal R}$. The metric field equations turn out be non-linear and they are beautifully expressed in terms of the Einstein tensor $G_{\mu\nu}$=${\kappa}T_{\mu\nu}$, where $\kappa$ denotes the coupling and $T_{\mu\nu}$ ensures the energy-momentum tensor. Importantly the $T_{\mu\nu}$ is believed to source the intrinsic curvatures of space-time. There were numerous attempts in the past hundred years to apprehend the $T_{\mu\nu}$ which could describe the geometry in $G_{\mu\nu}$ but none of them were completely successful \cite{ashtekar1979energy,o1986total, sharif2004energy,gumrukccuouglu2019role}. In fact, a firm believe leading to a conjecture seeking an appropriate $T_{\mu\nu}$ remains unanswered. 

\sp
\no
The unresolved issue in Einstein gravity provokes thought to believe in an alternate formulation \cite{wilczek1998riemann, blagojevic2013exotic,anastasiou2018gravity,benisty2022dark} to go beyond General Relativity (GR). This was with a plausible aim to find a satisfactory $T_{\mu\nu}$. Along the line, higher dimensional theories underlying the Riemann curvature tensor have been explored in the past \cite{cho1975higher, troncoso2000higher, wesson2015status}. In this paper we attempt to address an apparently unresolved problem in GR but with a higher dimensional gauge theoretic framework \cite{singh2013discrete,singh2013emergent,singh2014quantum}. Research ensures that an order two anti-symmetric $B_{\mu\nu}$ field underlying a $U(1)$ gauge symmetry in $(4$$+$$1)$ may serve as an insightful tool to find the required $T_{\mu\nu}$. Interestingly a geometric torsion (GT) theory is known to ensure an open geometry which is unlike to the closed geometries in GR. The observed perihelion precision \cite{nitish2020perspectives} reconfirms the open geometry in presence of an extra dimension to GR. Interestingly the proposed bulk/boundary correspondence resembles to an established  open/closed string duality \cite{sen2003open}.

\sp
\no
 In this article we revisit a GT description underlying a $5D$ bulk gauge theory with a renewed interest for a boundary gravity \cite{nitish2019cft6,gupta2020aspects}. A topological $BF$ term is argued to incorporate the winding modes in gravity. It is believed to ensure an UV finite quantum correction to GR. It was shown that a torsion connection modifies a covariant derivative $\nabla_{\mu}$$\rightarrow$$ {\cal D}_{\mu}$ in a $U(1)$ gauge theory \cite{singh2013discrete,nitish2021geometric}. The non-Riemann curvature tensors were derived in the modified theory which have been checked to ensure the space-time beyond GR. It was demonstrated that an emergent metric underlying the known black hole geometries in Einstein gravity is a low energy phenomenon derived from a symmetric matrix comprising the quantum fluctuations of $B_{\mu\nu}$ field. Thus the modified gauge theory with the Minkowski signature of space-time has been shown to incorporate intrinsic curvatures in presence of a GT connection. 

\sp
\no
It was argued that a gravitational pair of membrane/anti-membrane across the horizon of a background black hole can be created by the quantum of a GT in a gauge theory. This is analogous to the pair creation process ($\gamma\rightarrow e^+e^-$) known in quantum electrodynamics. Similarly the $B_{\mu\nu}$ field quantum is believed to vacuum create a pair of string/anti-string \cite{bachas1992pair}. In fact a vacuum pair creation is a powerful quantum tool and is believed to provide an enhanced vision to explore more in the quantum gravity domain. Interestingly the idea has been exploited to explain the Hawking radiation phenomenon at the horizon of a black hole \cite{hawking1975particle}. We may mention that the tool was also beautifully explored to address a cosmic pair 
\cite{ majumdar2002cosmological}. 

\section{\label{sec:level2}Aspects of multi black holes in GR}
Reissner-Nordstr\"{o}m (RN) black hole is known as exact in GR. It is characterised by two unequal deformation parameters ($M$$>$$|Q_e|$) and they source two independent potentials within the RN black hole. The instability with two horizons in RN black hole stabilizes to an extremal geometry with an equipotential. Thus a charged, or more than one parameter, black hole is likely to provide a clue to unfold some aspects of quantum gravity. Interestingly a drastic change in feature from an extremal RN black hole to multi black holes in isotropic coordinates \cite{carroll2019spacetime} may be perceived with a notion of cloning in a  quantum scenario. In fact, the cloning of black hole is believed to be associated with the non-locality on the event horizon \cite{singh2014quantum}. Intuitively a non-local horizon in an extremal RN black hole may be viewed as a collection of closely spaced horizons which ensure the exact multi black holes in GR. 

\vspace{.1in}
\noindent 
In addition, an extremal black hole is believed to provide a clue to a quantum tunneling phenomenon between the multiple black holes. Intuitively, the non-zero integer values of the conserved charge $M$=$|Q_e|$ reconfirms the multiple black holes. We recall that $M$ takes continuum range of values and a typical electric charge can take non-zero integer values in a non-extremal RN black hole. However the RN black hole shrinks ($r_+\leftrightarrow r_-$) to ensure a stable configuration for a positive integer value to $M$. This in turn would describe an extremal RN multi black hole as $M$ takes $M_i$ positive integer values. A similar analysis in an emergent quantum gravity scenario \cite{singh2013emergent} would like to incorporate the multi black hole scenario even for non-extremal RN black hole. We postpone a detailed analysis to the section IV. 

\section{\label{sec:citeref}Boundary Quantum Gravity}
A promising theoretical idea relating a bulk $B_{\mu\nu}$ field theory to a boundary gravity \cite{gupta2020aspects} is believed to be insightful to a theory of quantum gravity. The boundary action comprises a local sector GR and a global sector which turns out to be topological. In the case the holographic correspondence naturally ensures that an emergent gravity is described by a closed theory and the bulk is an open theory. At this point we may recall the established $AdS_5$ bulk/boundary $CFT_4$ correspondence \cite{maldacena1999large}, where the gravity in bulk describes an isolated system due to the $AdS$ radius. However this feature is not feasible with an arbitrary geometry in bulk. 

\sp
\no
Now we begin with a $B_{\mu\nu}$ field dynamics ensuring a $U(1)$ flux $H_{\mu\nu\lambda}$=$3\nabla_{[\mu}B_{\nu\lambda]}$ in a perturbation gauge theory. A covariant constant ${\tilde B}_{\mu\nu}$ field is considered in addition to the $B_{\mu\nu}$ field dynamics. Interestingly a constant ${\tilde B}_{\mu\nu}$ has been explored as a perturbation parameter. Remarkably ${\tilde B}_{\mu\nu}$ theory has been shown to incorporate higher order perturbation corrections in ${\tilde B}_{\mu\nu}$ field. In fact the corrections were vital to envisage the gravitational interactions in a plausible quantum theory. An iteration replaces $H_{\mu\nu\lambda} \rightarrow {\cal H}_{\mu\nu\lambda}$. Then a term in GT is given by 
\begin{equation}
{\cal D}_\lambda {\tilde B}_{\mu\nu}= \nabla_\lambda {\tilde B}_{\mu\nu}+\frac{1}{2}{\cal H}_{\lambda\mu}{}^{\rho}{\tilde B}_{\rho\nu}-\frac{1}{2}{\cal H}_{\lambda\nu}{}^{\rho}{\tilde B}_{\rho\mu}\label{Mod-2}
\end{equation}  
Then the GT flux is given by
\begin{eqnarray}
{\cal H}_{\mu \nu \lambda}&=& H_{\mu\nu\lambda}+H_{\mu \nu}{}^{\alpha}{\tilde B}_{\alpha \lambda}+H_{\nu \lambda}{}^{\alpha} {\tilde B}_{\alpha \mu}\nonumber\\
&&\qquad\qquad\qquad\quad +H_{\lambda \mu}{}^{\alpha} {\tilde B}_{\alpha \nu} + {\cal O}({\tilde B}^2)\label{Mod-33}
\end{eqnarray}
With an appropriate coupling or length scale $l$ the $U(1)$ gauge transformation becomes
$\delta {\tilde B}_{\mu\nu}$=$l(\partial_{\mu}\Lambda_{\nu}$$-$$\partial_{\nu}\Lambda_{\mu})$. Explicitly a variation of the action yields
\begin{eqnarray}
\delta {\cal H}_{\mu \nu \lambda}^{2} 
=6l\left({\cal H}^{\mu\nu\lambda}H_{\mu\nu}{}^{\alpha}-{\cal H}^{\mu\nu\alpha}H_{\mu\nu}{}^{\lambda}\right)\nabla_{\alpha}\Lambda_{\lambda}\label{M0d-34}
\end{eqnarray}
The broken $U(1)$ gauge symmetry may be restored \cite{singh2013discrete} with a symmetric matrix $f_{\mu\nu}$$=$$\pm l^2 ( {\cal H}_{\mu\alpha\beta} H^{\alpha\beta}{}_{\nu})$.
It shows that a non-linear GT and a linear gauge theoretic torsion together ensure the matrix fluctuations and they have been argued to source an emergent quantum gravity potential $V_q$ in ref.\cite{nitish2020perspectives}. Analysis reveals that $V_q$ should be a non-Newtonian potential and hence is believed to be defined with a lower cutoff on the radial coordinate. Alternately, the required UV cutoff can also be governed by a non-point (or extended) charge. It was argued that a non-point charge, naturally modifies a point notion, and incorporates winding modes into the theory. In a large (length) scale limit, the symmetric matrix reduces to a non-trivial metric correction to the otherwise flat metric $g_{\mu\nu}$. It is given by 
\begin{equation}
{\tilde g}_{\mu\nu}=g_{\mu\nu}\pm f_{\mu\nu}\ |_{\rm large\ scale}\label{Mod-5}
\end{equation}
With the Minkowski signature of space-time and for on-shell ${\tilde B}_{\mu\nu}$, the generic curvature tensor has been shown to describe a $4$-flux ${\cal F}_{\mu\nu\lambda\rho}$=$4l\nabla_{[\mu}{\cal H}_{\nu\lambda\rho ]}$ in addition to ${\cal K}_{\mu\nu\lambda\rho}$. The reduced effective curvatures in the case are given by
\begin{eqnarray}
&&\left[{\cal D}_{\mu}, {\cal D}_{\nu}\right] A_{\lambda}=\left ({\cal K}_{\mu\nu\lambda}{ }^{\rho}+{\cal F}_{\mu\nu\lambda}{ }^{\rho}\right )A_{\rho}\nonumber\\
&&\qquad\qquad\qquad\quad  +\left ({\cal F}_{\nu\mu}
+{\cal H}_{\mu\nu}{}^{\rho} {\cal D}_{\rho}\right ) A_{\lambda}\ ,\nonumber\\
{\rm where}\quad &&{{\cal K}}_{\mu \nu \lambda}{}^{\rho}=\frac{1}{4}\left({\cal H}_{\mu \lambda}{}^{\sigma} {\cal H}_{\nu \sigma}{}^{\rho}-{\cal H}_{\nu \lambda}{ }^{\sigma} {\cal H}_{\mu \sigma}{}^{\rho}\right)
\label{Mod-9}
\end{eqnarray}
Then the effective action \cite{gupta2020aspects} underlying an open geometry (say $R\times S_4$ topology) becomes
\begin{equation}
S_{\rm V}= 
\frac{-1}{48l^3}\int d^5x\sqrt{-g}\ \left (l^2{\cal F}_{\mu\nu\lambda\rho}^2 +4 {\cal H}_{\mu\nu\lambda}^2 + 4H_{\mu\nu\lambda}^2\right )\label{generic-bulk5} 
\end{equation}
A priori, the first term reconfirms a non-interacting GT in bulk with one propagating degree of freedom (PDF). It can be checked to confirm a dynamical axion in the quantum theory. An axion is known to source an instanton and is believed to ensure the quantum tunneling among the multiple vacua. The second term in eq(\ref{generic-bulk5}) ensures a non-canonical interaction in the perturbation theory. The third term describes a free field dynamics. Altogether the generic bulk dynamics is governed by two interacting fields (GT and ${\tilde B}_{\mu\nu}$) and a non-interacting $B_{\mu\nu}$ in the  perturbation theory. The background ${\tilde B}_{\mu\nu}$ independence of the flux $H_{\mu\nu\lambda}$ allows the exterior calculus to re-express the torsion as a $3$-form $H_3$$=$$dB_2$. The third term in eq(\ref{generic-bulk5}) is a total derivative and hence ensures a boundary term $B_2$$\wedge$$F_2$. In the case the open geometry with a bulk GT dynamics associates a boundary $S_4$ with a propagating axion along $R$. The vacuum expectation $<$$\chi$$>$$=$$\chi_0$$=$ (cosmological) constant $\Lambda$ ensures that the boundary space-time curvature is primarily sourced by an interacting second term in eq(\ref{generic-bulk5}). 

\sp
\no
 Remarkably the precise curvature on the boundary theory has been argued to map to ${\cal R}$ under a proposed correspondence between the bulk GT and boundary GR \cite{nitish2019cft6, gupta2020aspects}. In fact the bulk/boundary gravity correspondence was primarily based on two key evidences and they are: (i) ${\cal K}_{\mu\nu\lambda\rho}$ shares both the symmetry properties of ${\cal R}_{\mu\nu\lambda\rho}$ under the interchange of Lorentz indices independently and in pairs though they are respectively sourced by  $B_{\mu\nu}$ field and metric field \cite{singh2013discrete} and (ii) generically the PDF of a $B_{\mu\nu}$ field  in bulk is always one degree higher than that of a metric field on the boundary \cite{gupta2020aspects}. Thus with a bulk/boundary correspondence, the second term ensures a Ricci like scalar ${\cal K}$ and may be identified with ${\cal R}$. With a subtlety the boundary action becomes
\begin{equation}
S_{\partial V}=\frac{1}{2\kappa}\int d^4x\sqrt{-g}\ \left ( 
{\cal R}-2\Lambda\right ) - \int B_2\wedge F_2
\label{boundary} 
\end{equation}
It reconfirms that the boundary (quantum) theory is generically described by GR with a topological correction. In this article we show that the boundary dynamics in the emergent scenario is an exact. A nonzero topological invariant in GR such as Euler characteristics $\chi$ or signature $\tau$ is believed to unfold the presence of winding modes. However an extremal RN ensures $\chi$=$0$=$\tau$ and the quantum correction is apparently decoded with a non-local horizon. This in turn describes multi black holes as an exact in GR. At a first sight, these two alternate descriptions are believed to be in agreement with the non-unique nature of quantum gravity.

\section{RN type multi black holes}
We begin with the gauge ansatz in $(3$$+$$1)$ dimensions \cite{singh2013emergent} for $(b,P)>0$. For simplicity we consider $l$$=$$1$ and then the ansatz becomes
\begin{eqnarray}
&&B_{t\theta}=B_{r\theta}=b\; ,\quad {\tilde B}_{\theta\phi}=p\ 
\sin^2\theta\ ,\nonumber\\ 
&&A_t= {{Q_e}\over{r}}\; ,\quad A_\phi=-Q_m\cos\theta
\end{eqnarray}
They have been used to obtain the line-element for a RN type geometry in a large length scale limit. Then the line element becomes 
\begin{eqnarray}\label{em RNBH}
ds^{2}&=&-\left(1-\frac{b^{2}}{r^{2}} + \frac{Q_{e}^{2}}{r^{4}}\right)dt^{2}+\left(1-\frac{b^{2}}{r^{2}}+ \frac{Q_{e}^2}{r^{4}}\right)^{-1}dr^{2}\nonumber\\
&&\qquad\qquad\qquad\qquad\quad\ \ +\ r^{2}\left(1-\frac{Q^2_m}{r^4} \right) d \Omega^{2}\ ,
\end{eqnarray} 
where the integers $(b,Q_e,Q_m)$ are the conserved charges. 
The emergent line-element empirically confirms a $(3$$+$$1)$ dimensional dyonic black hole, with a mass $b^2$, embedded in $(4$$+$$1)$. Since the contribution to geometry from the mass, electric and magnetic charges take on positive numbers $(1,4,9,16 \dots)$, their spectrum describes bound states of non-extremal black holes for $b^2$$>$$\pm 2Q_e^2$ and with horizon radii $r_{\pm}$ satisfying $2r_{\pm}^{2}$$=$$b^{2}$$\pm$$ \sqrt{b^{4}-4 Q^{2}_e}$. However each bound state describes a continuum space-time within and is separated from all other bound states by classically forbidden regimes. Nevertheless the quantum gravity scenario allows tunneling between the black holes. It is a new feature associated with the non-extremal RN type black hole which is unlike to that in GR. 

\sp
\no
In addition the multi RN type black holes are defined with a reduced radius of $S_2$ due to the magnetic charge. The $S_2$ area becomes singular at $r^4$$=$$Q_m^2$. With an imposed self duality and in a limit $r^4$$\rightarrow$$Q_e^2$, the emergent RN identifies with the causal patch of a Schwarzschild black hole with a reduced effective mass $(b^2/2)$ embedded in a higher dimension. This is due to a fact that the action, for a $U(1)$ gauge field, vanishes for a self dual electromagnetic field which ensures a vacuum solution. Interestingly some of the emerging features share with that of a $(3$$+$$1)$ dimensional charged black hole in string theory \cite{garfinkle1991charged}.

\sp
\no
Since the causal patch is independent of the magnetic charge, one may set $Q_m$$=$$0$ in the extremal RN type black hole without any loss of generality. With a shift in the square of the radial coordinate  $\rho^{2}$=$(r^{2}$$-$$Q_{e})$ the line element in isotropic coordinates become
\begin{eqnarray}
ds^{2}&=&-H^{-2} d t^{2}+H\left (d { \rho}^{2}+{ \rho}^{2} d \Omega^{2}\right ),\nonumber\\ 
{\rm where}\quad H&=&\left (1+ {{Q_{e}}\over{\rho^{2}}}\right )
\label{extbh}
\end{eqnarray}
It is straight-forward to check that a time independent $H$ satisfies the Laplace's equation
$\nabla^{2} H$=$0$ in $(4$$+$$1)$ dimensions. Its solution is given by
\begin{eqnarray}
H=1+\sum_{i=1}^{N} \frac{m_i}{\left|y-y_{i}\right|^2}
\end{eqnarray}
The harmonic function $H$ is well behaved at the spatial infinity. Each value of $i$ defines a horizon and hence the solution (\ref{extbh}) describes an extremal RN multi black hole in near horizon geometry even in an emergent gravity scenario. It provokes one to believe in a nonlocal horizon which in turn is a collection of a multi horizons along a radial coordinate $\rho$. In fact an extremal RN type is not derived as an exact solution in GR. The multi black holes are obtained in the limit for a large scale structure of space-time in a quantum gravity theory. 

\section{Boundary GR from the Bulk $T_{\mu\nu}$}
In this section we would like to review the validity of the conjectured GT in bulk/boundary gravity correspondence. In particular we check if the Einstein tensor type in an emergent scenario is an exact in GR? 
The non-vanishing components of the Einstein type tensor ${\tilde G}_{\mu\nu}$$=$$({\tilde{\cal R}}_{\mu\nu}$$-$$\frac{1}{2}{\tilde g}_{\mu\nu}{\tilde{\cal R}})$ are worked out with near horizon coordinate to yield
\begin{eqnarray}
{\tilde G}_{\theta \theta}&=&\frac{2 Q_e\left(2 Q_e-{ \rho}^{2}\right)}{\left({ \rho}^{2}+Q_e\right)^{2}}\ ,\;\  {\tilde G}_{\phi \phi}=\frac{2 Q_e\left(2 Q_e-{ \rho}^{2}\right)}{\left({ \rho}^{2}+Q_e\right)^{2}} \sin ^{2} \theta\nonumber\\
{\tilde G}_{tt }&=&\frac{{ \rho}^{4} Q_e\left(Q_e-2 { \rho}^{2}\right)}{\left({ \rho}^{2}+Q_e\right)^{5}}\; ,\; 
{\tilde G}_{rr}=-\frac{Q_e\left(Q_e-2 { \rho}^{2}\right)}{{ \rho}^{2}\left({ \rho}^{2}+Q_e\right)^{2}}\label{bulk-1}
\end{eqnarray}
The components are re-expressed in term of the radial coordinate and they become
\begin{eqnarray}
{\tilde G}_{tt}&=&\frac{3 Q_e^{4}}{r^{10}}-\frac{8 Q_e^{3}}{r^{8}}+\frac{7 Q_e^{2}}{r^{6}}-\frac{2 Q_e}{r^{4}}\nonumber\\
{\tilde G}_{rr}&=&\frac{-Q_e^{2}}{\left(r^{2}-Q_e\right) r^{4}}+\frac{2 Q_e}{r^{4}}\nonumber\\
{\tilde G}_{\theta\theta}&=&\frac{6 Q_e^{2}}{r^{4}}-\frac{2 Q_e}{r^{2}}\nonumber\\ 
{\tilde G}_{\phi\phi}&=&\frac{6 Q_e^{2} \sin ^{2} \theta}{r^{4}}-\frac{2 Q_e \sin ^{2} \theta}{r^{2}}
\end{eqnarray}
It is important to note that the conserved charges on a brane is opposite in sign to that on an anti-brane \cite{singh2013discrete}. The components of Einstein type tensor on a vacuum created gravitational pair of $(3{\bar 3})$-brane are worked out to yield
\begin{eqnarray}
{\tilde G}_{tt}&=&\frac{6 Q_e^{4}}{r^{10}}+\frac{14 Q_e^{2}}{r^{6}}\ ,\quad 
{\tilde G}_{rr}=-\frac{2 Q_e^{4}}{r^{10}}-\frac{2 Q_e^{2}}{r^{6}}\nonumber\\
{\tilde G}_{\theta\theta}&=&\frac{12 Q_e^{2}}{r^{4}}\qquad {\rm and}\quad\ {\tilde G}_{\phi\phi}=\frac{12 Q_e^{2} \sin ^{2} \theta}{r^{4}}
\end{eqnarray}
Now it is interesting to check the validity of Einstein like tensor for the exactness $i.e.\ {\tilde G}_{\mu\nu}$=${\tilde\kappa} T_{\mu\nu}$, where ${\tilde\kappa}$ is believed to be an effective gravitational coupling. Alternately, at least the 
${\tilde G}_{\mu\nu}$ should be proportional to $T_{\mu\nu}$ if the emergent gravity scenario corresponds to the Einstein gravity. In fact we would like to examine the above mentioned conjecture by explicitly computing the components of $T_{\mu\nu}$ derived from a bulk gauge theory in the case. It is important to realize that the perturbation gauge theory in the case incorporates the dynamics of ${\tilde B}_{\mu\nu}$ as well as the three index GT which respectively sources $T_{\mu\nu}(1)$ and $T_{\mu\nu}(2)$. Together they define the total $T_{\mu\nu}= T_{\mu\nu}(1)+T_{\mu\nu}(2)$. Explicitly they are given by
\begin{eqnarray}
T_{\mu\nu}(1)&=&{\cal H}_{\mu\alpha\beta}{\cal H}^{\alpha\beta}{}_{\nu}-\frac{g_{\mu\nu}}{6}{\cal H}_{\alpha\beta\gamma}{\cal H}^{\alpha\beta\gamma} \nonumber \\
\qquad {T}_{\mu\nu}(2)&=&{\cal F}_{\mu\alpha\beta\gamma}{\cal F}^{\alpha\beta\gamma}{}_{\nu}-\frac{g_{\mu\nu}}{8}{\cal F}_{\alpha\beta\gamma\delta}{\cal F}^{\alpha\beta\gamma\delta}
\end{eqnarray}
Invoking the gauge ansatz \cite{singh2013discrete} we write 
\begin{eqnarray}
B_{t\psi}&=&B_{r\psi}=b\; ,\quad  
B_{\theta\psi}=\tilde{P}^3 \sin^2\psi\cot\theta\nonumber\\
{\rm and}\quad {\tilde B}_{\psi\phi}&=&{P^3}\sin^2\psi\cos\theta
\end{eqnarray}
The non-vanishing components of GT are
\begin{eqnarray}
{\cal H}_{\theta\phi}{}^{\psi}&=& \frac{P^3}{r^2}\sin^2\psi \sin\theta \\
{\rm and }\quad {\cal H}_{\theta\phi}{}^{t}&=&-{\cal H}_{\theta\phi}{}^{r}=\frac{bP^3}{r^2} \sin^2\psi \sin\theta \label{geometric torsion}
\end{eqnarray}
Subsequently the nontrivial components of the $4$-index flux \cite{gupta2020aspects} become
\begin{eqnarray}
{\cal F}_{t\psi\theta\phi}&=&-{\cal F}_{r\psi\theta\phi}=\frac{-2bP^3}{r^2}\sin{2}\psi \sin\phi\nonumber\\
{\cal F}_{tr\theta\phi}&=&\frac{2bP^3}{r^3}\sin^2\psi\sin\theta\label{4form}
\end{eqnarray}
Explicitly the components of both the energy-momentum tensors are given by
\begin{eqnarray}
T_{tt}(1) &=&\left(1+\frac{2 b^{2}}{r^{2}}\right) \frac{P^{6}}{r^{6}}\; ,\quad
T_{rr}(1)=\left(-1+\frac{2 b^{2}}{r^{2}}\right) \frac{P^{6}}{r^{6}}\nonumber\\
T_{ \psi \psi}(1)&=&\frac{P^{6}}{r^{4}}\ ,\ T_{\theta\theta}(1)=\frac{P^{6}}{r^{4}}\sin^2\psi\nonumber\\
T_{\phi\phi}(1)&=&\frac{P^{6}}{r^{4}}\sin^2\psi\sin^2\theta\nonumber\\
{T}_{tt}(2)&=&-\frac{12{b}^2{P}^6}{r^{10}}\left(8\cot^2\psi+3\right)\nonumber\\
{T}_{rr}(2)&=&\frac{12{b}^2{P}^6}{r^{10}}\left(-8\cot^2\psi+3\right)
\nonumber\\ 
{T}_{\psi\psi}(2)&=&\frac{12{b}^2{P}^6}{r^{8}}\ ,\; 
{T}_{\theta\theta}(2)=\frac{36{b}^2{P}^6}{r^{8}}\sin^2\psi\nonumber\\
{T}_{\phi\phi}(2)&=&\frac{36{b}^2{P}^6}{r^{8}}\sin^2\psi\sin^2\theta
\end{eqnarray}
 We consistently fix an angular coordinate $\psi$=${{\pi}\over{2}}$ to arrive at a $(3$$+$$1)$-dimensional embedded regime within a $(4$$+$$1)$ bulk. Then the energy-momentum tensors in a specific combination of their components have been checked to satisfy the field equations. A priori they are defined with eight constants 
$({\kappa}_i, {\kappa}_i)$ for $i$=$(0,1,2,3)$ and the field equations are given by 
\begin{eqnarray}
{\tilde G}_{tt}&=&\kappa_0 T_{tt}(1)-{\tilde\kappa}_0T_{tt}(2)-{\tilde\kappa}_3 T_{\phi\phi}(2) \nonumber\\
{\tilde G}_{rr}&=&\kappa_1 T_{rr}(1)-{\tilde\kappa}_1 T_{rr}(2)-{\tilde\kappa}_2 T_{\theta\theta}(2) \nonumber\\
{\tilde G}_{\theta\theta}&=&\kappa_2 T_{\theta\theta}(1) \quad 
{\rm and}\quad {\tilde G}_{\phi\phi}=\kappa_3 T_{\phi\phi}(1)\label{FieldEQ}
\end{eqnarray}
The emergent equations take the usual form of Einstein field equations.
The conditions are worked out to yield 
\begin{eqnarray}
\kappa_0&=&7\kappa_1=\frac{7}{6}\kappa_2=\frac{7}{6}\kappa_3 = 126{\tilde{\kappa}}_2=18{\tilde{\kappa}}_3\sin^2\theta\nonumber \\
{\tilde{\kappa}}_0&=&\frac{1}{6}\quad {\rm and} \quad{\tilde{\kappa}}_1=\frac{1}{18} 
\end{eqnarray}
They ensure only one independent constant $\kappa$ among the eight $(\kappa_i, {\tilde\kappa}_i)$. All the resulting field equations (\ref{FieldEQ}) can be checked to be defined with a nontrivial coupling as is believed in GR. The result reconfirms our conjecture that the RN type in an emergent gravity is an exact in GR. Analysis presumably identifies an emergent gravity scenario with that of GR at least for the RN black hole. 

\sp
\no
In addition the complete $T_{\mu\nu}$ has been shown to be sourced by two independent fields in a perturbation theory. Importantly a combination of energy-momentum tensor components for both the fields were shown to source the Riemann curvature for the extremal RN black hole. The result is remarkable and is believed to strengthen our proposed correspondence \cite{nitish2019cft6} between a gauge theory in bulk and a boundary quantum gravity underlying the Einstein gravity.  
   
\section{Concluding remarks}
In this article we have revisited a correspondence between the $5D$ bulk $U(1)$ gauge theory and a boundary gravity\cite{gupta2020aspects} with a renewed interest. It was shown that the boundary gravity dynamics comprises GR and a topological $BF$ term which in turn describes a mass dipole correction to GR. It is argued that the topological correction incorporates winding mode(s) into the otherwise point particle notion in GR. Arguably the winding modes ensure an UV finite (non-perturbation) quantum correction to GR. 

\sp
\no
We have shown that the extremal RN type multi black holes are exact in GR. In particular the energy-momentum tensor in the boundary gravity was shown to be a special case of the bulk dynamics ensured by two completely anti-symmetric fluxes ${\cal H}_{\mu\nu\lambda}$ and ${\cal F}_{\mu\nu\lambda\rho}$ in a perturbation theory. Analysis reveals that the dark energy, in the disguise of a bulk axion, and a GT together ensure the required $T_{\mu\nu}$ to source the geometry in GR.  

\sp
\no 
{\bf Acknowledgements:} Author (SK) gratefully acknowledges the research grant-in-aid 2021-22 from the Institute of Eminence at the University of Delhi.



\end{document}